\RequirePackage{filecontents}

\RequirePackage{snapshot}
\documentclass[aps,prd,reprint,showpacs,secnumarabic,groupedaddress,amsmath,amssymb]{preprint}

\renewcommand\documentclass[2][]{}
\let\origparagraph=\paragraph
\AtBeginDocument{%
  \let\paragraph=\origparagraph
}
\makeatother
\documentclass[%
aps,
pra,
twocolumn,
groupedaddress,
showpacs,
floatfix,
amsmath,
amssymb,
]{revtex4-1}
\pdfoutput=1

\usepackage{graphicx}

\graphicspath{{figs/}{summary/}}

\usepackage[utf8]{inputenc}
\usepackage[T1]{fontenc}
\usepackage{txfonts}
\usepackage{tgtermes}
\usepackage[altg]{qtxmath}

\usepackage{color}
\usepackage{bm}
\usepackage{braket}
\usepackage{microtype}

\usepackage[%
pdfborder={0 0 0}, 
breaklinks=true,
pdfauthor={Anton W{\"o}llert, Heiko Bauke, Christoph H. Keitel},
pdftitle={Spin polarized electron-positron pair production via elliptical polarized laser fields}]
{hyperref}
\definecolor{black}{RGB}{0,0,0}
\definecolor{darkblue}{RGB}{0,0,90}
\definecolor{lightgray}{RGB}{80,80,80}
\hypersetup{
  colorlinks=true,
  linkcolor=darkblue,
  anchorcolor=darkblue,
  citecolor=darkblue,
  filecolor=darkblue,
  menucolor=darkblue,
  runcolor=darkblue,
  urlcolor=darkblue,
}

\newcommand{\scalprod}[2]{\langle #1 \vert #2 \rangle}

\newcommand{\norm}[1]{\vert #1 \vert}
\newcommand{\op}[1]{\hat{#1}}
\renewcommand*{\vec}[1]{\bm{#1}}
\newcommand*{\D}{\mathrm{d}}
\newcommand*{\I}{\mathrm{i}}
\newcommand*{\e}{\mathrm{e}}
\renewcommand*{\Re}[1]{\operatorname{Re}{#1}}

\newcommand{\tin}{t_{\mathrm{in}}}
\newcommand{\tout}{t_{\mathrm{out}}}

\newcommand*{\trans}[1]{#1^{\mathsf{T}}}
\newcommand*{\herm}[1]{#1^{\dagger}}
\newcommand*{\HD}{\mathcal{H}}
\newcommand*{\phiin}[1][\pm]{{}_{#1}\varphi}
\newcommand*{\phiout}[1][\pm]{{}^{#1}\varphi}
\newcommand*{\Ein}[1][\pm]{{}_{#1}\varepsilon}
\newcommand*{\Eout}[1][\pm]{{}^{#1}\varepsilon}
\newcommand*{\Gud}[2]{G({}^#1|{}_#2)}

\allowdisplaybreaks

\begin{document}

\title{Spin polarized electron-positron pair production via elliptical
  polarized laser fields}

\author{Anton W{\"o}llert}%
\email{woellert@mpi-hd.mpg.de}
\affiliation{Max-Planck-Institut f{\"u}r Kernphysik, Saupfercheckweg 1, 69117 Heidelberg, Germany}

\author{Heiko Bauke}
\email{heiko.bauke@mpi-hd.mpg.de}
\affiliation{Max-Planck-Institut f{\"u}r Kernphysik, Saupfercheckweg 1, 69117 Heidelberg, Germany}

\author{Christoph H.~Keitel}
\affiliation{Max-Planck-Institut f{\"u}r Kernphysik, Saupfercheckweg 1, 69117 Heidelberg, Germany}

\date{\today}

\begin{abstract}
  We study nonperturbative multiphoton electron-positron pair creation
  in ultrastrong electromagnetic fields formed by two
  counterpropagating pulses with elliptic polarization.  Our numerical
  approach allows us to take into account the temporal as well as the
  spatial variation of the standing electromagnetic field.  The spin and
  momentum resolved pair creation probabilities feature characteristic
  Rabi oscillations and resonance spectra.  Therefore, each laser
  frequency features a specific momentum distribution of the created
  particles.  We find that, depending on the relative polarization of
  both pulses, the created electrons may be spin polarized along the
  direction of field propagation.
\end{abstract}


\maketitle

\section{Introduction}
\label{sec:intro}

Quantum electrodynamics predicts the possible breakdown of the vacuum in
the presence of ultrastrong electromagnetic fields into pairs of
electrons and positrons.  In the seminal articles by Sauter and others
\cite{sauter1931verhalten,heisenberg1936folgerungen,schwinger1951on_gauge},
pair creation was investigated theoretically for constant
electromagnetic fields.  Since then this process has been studied in
numerous works; see
Refs.~\cite{ehlotzky2009fundamental,dipiazza2012extremely} for recent
reviews.  The Schwinger critical field strength of
$E_\mathrm{S}=1.3\times10^{18}\,\text{V/m}$, where spontaneous pair
creation is expected to set in, cannot be reached even by the strongest
laser facilities available today.  However, pair creation may be
assisted by additional fields or particles or by electromagnetic fields
that oscillate in time and space.  Novel light sources envisage to
provide field intensities in excess of $10^{20}\,\mathrm{W/cm^2}$ and
field frequencies in the \mbox{x-ray} domain
\cite{Altarell:etal:2007:XFEL,Yanovsky:Chvykov:etal:2008:Ultra-high_intensity-,McNeil:Thompson:2010:X-ray_free-electron,Emma:Akre:etal:2010:First_lasing,Mourou:Fisch:etal:2012:Exawatt-Zettawatt_pulse}.
The ELI-Ultra High Field Facility aims to reach intensities exceeding
even $10^{23}\,\mathrm{W/cm^2}$ corresponding to a field strength of
about $10^{15}\,\mathrm{V/m}$, which is only a few orders of magnitude
below the critical field strength $E_\mathrm{S}$.  It has been argued
that the intensity of such near-future ultrastrong light sources may be
sufficient to observe pair creation
\cite{Ringwald:2001:Pair_production,Alkofer:Hecht:etal:2001:Pair_Creation,Blaschke:Prozorkevich:etal:2006:Pair_Production,Bell:Kirk:2008:Possibility_of,Blaschke:Prozorkevich:etal:2009:Dynamical_Schwinger}.

Pair creation in laser fields is often studied by considering a
time-varying homogeneous electric field
\cite{brezin1970pair,Popov:1971:Production_of,Alkofer:Hecht:etal:2001:Pair_Creation,Roberts:Schmidt:etal:2002:Quantum_Effects,Blaschke:Prozorkevich:etal:2006:Pair_Production,Hebenstreit:Alkofer:etal:2009:Momentum_Signatures,Dumlu:Dunne:2010:Stokes_Phenomenon,mocken2010nonperturbative,Orthaber:Hebenstreit:etal:2011:Momentum_spectra}.
Neglecting the magnetic field component and the spatial variation of the
electric field eases the theoretical analysis and is motivated by the
fact that a standing electromagnetic wave, which is formed by
superimposing two counterpropagating linearly polarized waves of equal
wavelength, intensity, and polarization, has a vanishing magnetic field
component at positions, where the electric field reaches its maximum.  A
more realistic description of electron-positron pair creation, however,
requires the consideration of inhomogeneous electric fields
\cite{Nikishov:1970:Barrier_scattering,gies2005pair,Dunne:Wang:2006:Multidimensional_worldline,Kim:Page:2007:Improved_approximations,Kleinert:Ruffini:etal:2008:Electron-positron_pair}
or electromagnetic fields that depend on time and space
\cite{ruf2009pair,Bulanov:Mur:etal:2010:Multiple_Colliding,hebenstreit2011particle}.

Introducing the adiabaticity parameter $\xi=eE/(m_0c\omega)$
\cite{mocken2010nonperturbative}, where $m_0$ denotes the electron mass,
$c$ the speed of light, $e$ the elementary charge, $E$ the peak electric
field strength, and $\omega$ the electromagnetic field's angular
frequency, one can distinguish two limiting cases of pair creation in
oscillating electromagnetic fields \cite{brezin1970pair} with
frequencies $\omega\lesssim m_0c^2/\hbar$.  These two regimes are
commonly referred to as nonperturbative Schwinger pair creation
($\xi\gg1$) and perturbative multiphoton pair creation ($\xi\ll1$).
Note that these regimes are sometimes distinguished via the Keldysh
parameter of vacuum pair creation $\gamma$, which is the inverse of
$\xi$.  In this contribution we will focus on the intermediate regime of
nonperturbative multiphoton pair creation ($\xi\approx1$), which is most
challenging from the theoretical point of view and, therefore, less well
understood
\cite{ruf2009pair,mocken2010nonperturbative,Li:etal:2014:Mass_shift}.
In particular, we consider pair creation in a standing electromagnetic
wave formed by two counterpropagating waves.  This means we include not
only the electric field component but also the field's magnetic
component, which can alter the pair creation rate
\cite{Su:Su:etal:2012:Magnetic_Control,Wollert:Klaiber:etal:2015:Relativistic_tunneling}.
The temporal as well as spatial oscillations of these fields are
accounted for.  In our work the waves may have arbitrary elliptical
polarization in contrast to prior publications where mainly linear
polarization was considered.  Electron dynamics in intense light with
elliptical polarization
\cite{Bauke:Ahrens:etal:2014:Electron-spin_dynamics} and pair creation
via the nonlinear Bethe--Heitler process
\cite{Muller:Muller:2012:Longitudinal_spin,Muller:Hu:etal:2012:Electron_recombination}
can show nontrivial spin effects
\cite{Bauke:Ahrens:etal:2014:Electron-spin_dynamics,Muller:Muller:2012:Longitudinal_spin}.
Therefore, pair creation in elliptically polarized light beams is
expected to feature unexplored spin effects.

The manuscript is organized as follows.  In Sec.~\ref{sec:beams} we
describe the laser field configuration and pay special attention to the
polarization of the light. We lay down the theoretical foundation for
the numerical work in Sec.~\ref{sec:theory}.  The numerical methods are
explained in Sec.~\ref{sec:numerics}.  Our results are presented in
Sec.~\ref{sec:results}.  Typical Rabi oscillations of the pair creation
probability and properties of the pair creation resonances are
discussed.  These resonance spectra are investigated for different
polarizations and compared to a model that employs the dipole
approximation, i.\,e., neglecting the magnetic field component and the
spatial variation of the electric field.  The dependence of pair
creation on the final momenta of the created particles is discussed at
the end of Sec.~\ref{sec:results}.  Finally, we conclude in
Sec.~\ref{sec:conclude}.  For the reminder of this article, we will
employ units with $\hbar=1$ and $c=1$.\vspace{0ex plus 1ex}

\section{Elliptically polarized light beams}
\label{sec:beams}

In this work we will focus on a setup with two counterpropagating
elliptically polarized laser beams with equal wavelength~$\lambda$.
Note that two counterpropagating laser beams with different wavelengths
can always be transformed by a suitable Lorentz boost along the
propagation direction into the considered setup.  Furthermore, two
crossed laser beams with the same wavelength and arbitrary but nonzero
collision angle can always be transformed into the considered setup by a
suitable Lorentz boost perpendicularly to the propagation
direction. Similarly, two crossed laser beams with different wavelengths
and arbitrary but nonzero collision angle can always be transformed by a
suitable proper Lorentz transformation (which may include boosts as well
as rotations) into the considered setup. Thus, it is quite generic.

\begin{figure}
  \centering
  \includegraphics[width=\linewidth]{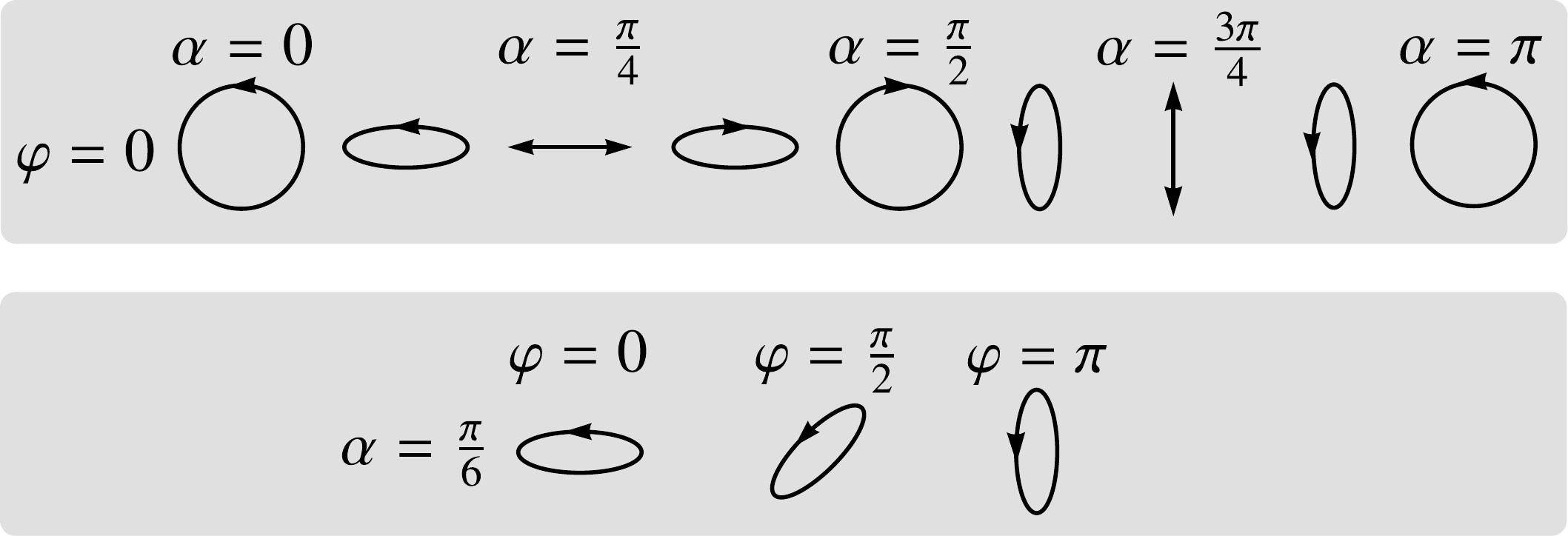}
  \caption{(color online) Schematic illustration of different
    polarization states as a function of the parameters $\alpha$ and
    $\varphi$, which determine the ellipticity (degree of polarization)
    and the orientation of the polarization ellipsis in the
    $x$-$y$~plane.  Upper row for constant orientation $\varphi$ but
    varying $\alpha$; lower row for constant ellipticity $\alpha$ but
    varying $\varphi$.}
  \label{fig:jones_vec}
\end{figure}

\begin{figure}
  \centering
  \includegraphics[width=0.9\linewidth]{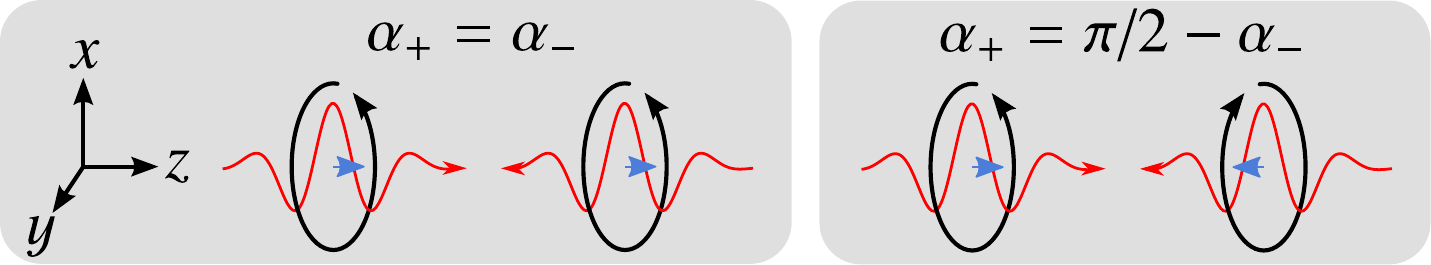}
  \caption{(color online) Schematic illustration of the possible
    relative spin orientations of two counterpropagating electromagnetic
    fields. Left: When the fields' ellipticity parameters equal
    ($\alpha_- = \alpha_+$), then both fields have the same sense of
    rotation.  Consequently, also the superposition of both fields
    rotates around the $z$ axis and this axis of rotation leads to a
    preferred direction along the positive or negative $z$ axis. Right:
    For $\alpha_- = \pi/2 - \alpha_+$ the sense of rotation is exactly
    opposite. Thus, the resulting field does not rotate, and no direction
    of the $z$~axis is preferred.}
  \label{fig:laser_spin}
\end{figure}

Let us denote the unit vectors along the coordinate axes of the
Cartesian coordinate system by $\vec{e}_x$, $\vec{e}_y$, and
$\vec{e}_z$.  We assume that the two light beams propagate along the
$z$~direction, and thus their wave vectors equal $\vec{k}_\pm = \pm k
\vec{e}_z$ with $k=2\pi/\lambda$.  The indices distinguish beams
traveling to the right (index ``$+$'') and to the left (index ``$-$''),
respectively. To describe the polarization of an electromagnetic wave,
it is convenient to use the Jones vector formalism
\cite{Jones:1941:New_Calculus}.  The so-called Jones vectors $\ket{l}$
and $\ket{r}$ for circular polarized beams with left-handed and
right-handed orientation are defined as
\begin{subequations}
  \begin{align}
    \ket{l} = & \frac{1}{\sqrt{2}} (\vec{e}_x +\I \vec{e}_y)\,,\\
    \ket{r} = & \frac{1}{\sqrt{2}} (\vec{e}_x - \I \vec{e}_y)\,.
  \end{align}
\end{subequations}
Any polarization state of the electric fields of the two laser beams can
be formed by a linear combination of the Jones vectors $\ket{l}$ and
$\ket{r}$.  At position $\vec r=(x,y,z)$ and time $t$, the electric field
of a single beam is given by
\begin{equation}
  \label{eqn:E_0_pm_complex}
  \vec{E}_{\pm}(\vec r, t) = \Re{\left(
      E \left( 
        \cos \alpha_\pm \ket{l} + \sin \alpha_\pm \, \e^{\I \varphi_\pm} \ket{r} 
      \right) \e^{\I(\pm k z - \omega t)} 
    \right)}\,,
\end{equation}
where $\omega=k$ denotes the angular frequency and $E$ the peak electric
field strength.  The expression \eqref{eqn:E_0_pm_complex} may be
transformed into
\begin{multline}
  \label{eqn:E_0_pm_real}
  \vec{E}_{\pm}(\vec r, t) =
  E
  \begin{pmatrix}
    \cos \varphi_\pm/2 & -\sin \varphi_\pm/2 \\
    \sin \varphi_\pm/2 & \phantom{-}\cos \varphi_\pm/2 \\
    0 & 0
  \end{pmatrix} \times \\
  \begin{pmatrix}
    \cos (\pm k z - \omega t + \varphi_\pm/2)\,\cos(\alpha_\pm - \pi/4) \\
    \sin (\pm k z - \omega t + \varphi_\pm/2)\,\sin(\alpha_\pm - \pi/4) \\
  \end{pmatrix} \,.
\end{multline}
The second factor in \eqref{eqn:E_0_pm_real} parametrizes an ellipse,
and the parameters $\alpha_+$ and $\alpha_-$ determine its ellipticity.
The parameters $\varphi_+$ and $\varphi_-$ specify the orientation of
this ellipse via the first factor in \eqref{eqn:E_0_pm_real}, which
represents a rotation matrix; see also Fig.~\ref{fig:jones_vec}.  The
light beams have linear polarization for $\alpha_\pm=\pi/4$ and
$\alpha_\pm=3\pi/4$, while they have circular polarization for
$\alpha_\pm=0$ and also for $\alpha_\pm=\pi/2$.  The magnetic field
component follows via
\mbox{$\vec{B}_{\pm}=\vec{k}_{\pm}\times\vec{E}_{\pm}/k$}, and the mean
intensity equals $\varepsilon_0E^2/2$ independent of $\alpha_\pm$ and
$\varphi_\pm$.  The corresponding Coulomb-gauge vector potential is
\begin{equation}
  \label{eqn:vector_pot}
  \vec{A}_{\pm}(\vec r, t) = \Re{\left(
      \frac{E}{\I\omega}
      \left( 
        \cos \alpha_\pm \ket{l} + \sin \alpha_\pm \, \e^{\I \varphi_\pm} \ket{r} 
      \right)
      \e^{\I(\pm k z - \omega t)}
    \right)} 
\end{equation}
\allowdisplaybreaks[0]%
or equivalently 
\begin{multline}
  \vec{A}_{\pm}(\vec r, t) = \frac{E}{\omega}
  \begin{pmatrix}
    \cos \varphi_\pm/2 & -\sin \varphi_\pm/2 \\
    \sin \varphi_\pm/2 & \phantom{-}\cos \varphi_\pm/2 \\
    0 & 0
  \end{pmatrix} \times \\
    \begin{pmatrix}
    \phantom{-}\sin (\pm k z - \omega t + \varphi_\pm/2)\,\cos(\alpha_\pm - \pi/4) \\
    -\cos (\pm k z - \omega t + \varphi_\pm/2)\,\sin(\alpha_\pm - \pi/4)
  \end{pmatrix} \,.
\end{multline}
The total vector potential of two counterpropagating electromagnetic
waves equals $\vec{A}=\vec{A}_{+}+\vec{A}_{-}$ and similarly for the
electric and magnetic field components.  In the reminder of this article,
we will assume that both counterpropagating light waves have the same
orientation, i.\,e., $\varphi_+=\varphi_-$, and that they have the same
ellipticity, i.\,e., $\alpha_-=\alpha_+$ or $\alpha_-=\pi/2-\alpha_+$
depending on the sense of rotation being identical or opposite to each
other; see Fig.~\ref{fig:laser_spin}.  Note that for any such field
configuration we can always find a coordinate system by a rotation
around the $z$~axis, where $\varphi_+=\varphi_-=0$.  Thus, we will set
in the following $\varphi_+=\varphi_-=0$ without loss of generality.
\allowdisplaybreaks

Elliptically polarized light beams carry photonic spin angular momentum
\cite{Bliokh:Bekshaev:etal:2013:Dual_electromagnetism}, which is
$\varepsilon_0E^2\cos(2\alpha_\pm)/(2\omega)$ for each of the two
considered counterpropagating light beams.  Thus, the ellipticity
paramaters $\alpha_+$ and $\alpha_-$ determine the photonic spin
density.  The total spin density of the considered setup equals
$2\varepsilon_0E^2\cos(2\alpha_\pm)/\omega$ (with orientation along the
$z$ direction) for the case that both individual beams have the opposite
helicity, i.\,e., $\alpha_+=\alpha_-$; see also the left part of
Fig.~\ref{fig:laser_spin}.  The total photonic spin density vanishes if
both light beams have same helicity, i.\,e., $\alpha_+=\pi/2-\alpha_-$;
see also the right part of Fig.~\ref{fig:laser_spin}.  In what follows, we
will name the field setup ``corotating'' if $\alpha_+ = \alpha_-$ and
``antirotating'' if $\alpha_+ = \pi/2 - \alpha_-$.

\section{Theoretical foundations}
\label{sec:theory}

Quantum electrodynamics has to be applied to describe pair creation in
electromagnetic fields \cite{gitman1991quantum}.  The Hamiltonian
describing the dynamics of a quantum field state in an external
electromagnetic field is given by
\begin{equation}
  \label{eqn:H_QED}
  \op{H}(t) = 
  \int\herm{\op{\psi}(\vec r)}\HD(t)\op{\psi}(\vec r)\,\D^3 r\,,
\end{equation}
where $\op{\psi}(\vec r)$ denotes the time-independent spinor field
operator and
\begin{equation}
  \HD(t) = \vec\alpha\cdot\big(-\I\vec\nabla-q\vec A(\vec r, t)\big) + m_0\beta 
\end{equation}
is the Dirac Hamiltonian for a particle with charge $q$ and rest mass
$m_0$.  Furthermore, $\vec\alpha=\trans{(\alpha_x,\alpha_y,\alpha_z)}$
and $\beta$ stand for the Dirac matrices
\cite{Gross:2004:Relativistic_Quantum,Thaller:2005:Advanced_Visual}.
Let us denote the times, when the interaction with the external
electromagnetic field sets in and ends, by $\tin$ and $\tout$,
respectively.  Thus, $\vec E(t)=0$ for $t\le\tin$, and for
$t\ge\tout$. The Dirac Hamiltonian may be diagonalized at times $\tin$
and $\tout$ by introducing the two complete sets of orthonormal
functions $\phiin_n(\vec r)$ and $\phiout_n(\vec r)$ \footnote{In
  general the function sets $\phiin_n(\vec r)$ and $\phiout_n(\vec r)$
  may not coincide because the vector potential may not vanish and be
  different at $\tin$ and $\tout$, depending on the gauge.},
\begin{subequations}
  \begin{align}
    \HD(\tin) \phiin_n(\vec r) & =
    \Ein_n \, \phiin_n(\vec r) \,,\\
    \HD(\tout) {}^{\pm} \varphi_n(\vec r) & = \Eout_n \,
    \phiout_n(\vec r) \,,
  \end{align}
\end{subequations}
where $\Ein[+]_n>0$, $\Ein[-]_n<0$ and $\Eout[+]_n>0$, $\Eout[-]_n<0$
denote the corresponding positive and negative eigenenergies, with $n$
labeling the quantum state.  Then the spinor field operator
$\op{\psi}(\vec r)$ may be decomposed in one of the two bases
\begin{subequations}
  \begin{align}
    \label{eqn:psi_in}
    \op{\psi}(\vec r) & = \sum_n 
    \phiin[+]_n(\vec r)\op{a}_n(\tin) + 
    \phiin[-]_n(\vec r)\herm{\op{b}}_n(\tin) \,,\\
    \label{eqn:psi_out}
    \op{\psi}(\vec r) & = \sum_n 
    \phiout[+]_n(\vec r)\op{a}_n(\tout) + 
    \phiout[-]_n(\vec r)\herm{\op{b}}_n(\tout) \,,
  \end{align}
\end{subequations}
where $\herm{\op{a}}$, $\op{a}$ and $\herm{\op{b}}$, $\op{b}$, denote
the creation and the annihilation operators for the electron and the
positron, respectively, at times $\tin$ and $\tout$.  Quantum field
states before and after interaction with definite numbers of particles
can always be written as products of the vacuum states $\ket{0, \tin}$
and $\ket{0, \tout}$ and a number of particle and antiparticle creation
operators,
\begin{subequations}
  \begin{align}
    \ket{\tin} & =
    \herm{\op{a}}_{n_1}(\tin)\herm{\op{a}}_{n_2}(\tin)\dots
    \herm{\op{b}}_{m_1}(\tin)\herm{\op{b}}_{m_2}(\tin)\dots
    \ket{0,\tin} \,,\\
    \ket{\tout} & =
    \herm{\op{a}}_{n_1}(\tout)\herm{\op{a}}_{n_2}(\tout)\dots
    \herm{\op{b}}_{m_1}(\tout)\herm{\op{b}}_{m_2}(\tout)\dots
    \ket{0,\tout} \,.
  \end{align}
\end{subequations}
The amplitude for the transition between these states is given by
\begin{equation}
  M_\mathrm{in\to out} = \braket{\tout | \op{U}(\tout, \tin) | \tin }\,,
\end{equation}
where $\op{U}(\tout, \tin)$ denotes the time evolution operator that
corresponds to the Hamiltonian $\op{H}(t)$ and maps the state
$\ket{\tin}$ at time $\tin$ to some later time $\tout$. 

After interaction the average number of electrons in quantum state $n$
is given by
\begin{equation}
  \mathcal{N}_n = \braket{
    \op{U}(\tout, \tin)\tin | 
    \herm{\op{a}_{n}}(\tout)\op{a}_{n}(\tout) | 
    \op{U}(\tout, \tin)\tin
  } 
\end{equation}
or equivalently
\begin{equation}
  \mathcal{N}_n = \braket{
    \tin | 
    \herm{\op{\tilde a}_{n}}(\tout)\op{\tilde a}_{n}(\tout) | 
    \tin
  } \,,
\end{equation}
where
\begin{equation}
  \op{\tilde a}_{n}(\tout) =
  \herm{\op{U}(\tout,\tin)}\op{a}_{n}(\tout)\op{U}(\tout, \tin)
\end{equation}
is the representation of the electron creation operator in the
Heisenberg picture with $\ket{\tin}$ as the reference state.  
In the Heisenberg picture, the spinor field operator $\op{\psi}(\vec r)$
is replaced by the time-dependent operator $\op{\tilde \psi}(\vec r, t)$
of which the evolution is determined by the Dirac equation
\begin{equation}
  \label{eqn:Dirac_eq}
  \I\dot{\op{\tilde \psi}}(\vec r, t) = \HD(t)\op{\tilde \psi}(\vec r, t) \,.
\end{equation}
At time $\tin$ the operator $\op{\psi}(\vec r, t)$ equals
\eqref{eqn:psi_in}, and at later times
\begin{equation}
  \op{\tilde \psi}(\vec r, t) =
  \herm{\op{U}(t,\tin)}\op{\psi}(\vec r)\op{U}(t, \tin) 
\end{equation}
or equivalently 
\begin{equation}
  \label{eqn:tilde_psi_G}
  \op{\tilde \psi}(\vec r, t) =
  \int G(\vec r, \tout; \vec r', \tin) \op{\psi}(\vec r)\,\D^3r'\,,
\end{equation}
where $G(\vec r, t; \vec r', t')$ denotes the complete Green function
of \eqref{eqn:Dirac_eq}.  Using
\begin{equation}
  \op{\tilde a}_{n}(\tout) = 
  \int \herm{\phiout[+]_n(\vec r)} \op{\tilde \psi}(\vec r, t) \,\D^3r\,,
\end{equation}
\eqref{eqn:psi_in}, and \eqref{eqn:tilde_psi_G}, we get
\begin{equation}
  \label{eqn:a_tilde}
  \op{\tilde a}_{n}(\tout) = 
  \sum_m \Gud++_{n;m}\op{a}_m(\tin) +  \Gud+-_{n;m}\herm{\op{b}}_m(\tin)
\end{equation}
with $\Gud\pm\pm_{n;m}$ defined as
\begin{equation}
  \Gud\pm\pm_{n;m} = 
  \iint \herm{\phiout_n(\vec r)} 
  G(\vec r, \tout; \vec r', \tin)
  \phiin_m(\vec r')\,\D^3r'\,\D^3r \,.
\end{equation}
Choosing the vacuum state $\ket{0, \tin}$ as the initial state
$\ket{\tin}$, one can show with \eqref{eqn:a_tilde} that
\cite{Gitman:1977:Processes_of}
\begin{equation}
  \label{eqn:cn}
  \mathcal{N}_n = 
  \sum_m \Gud+-_{n;m}^* \Gud+-_{n;m} =
  \sum_m \norm{\Gud+-_{n;m}}^2 \,,
\end{equation}
and therefore the expectation value of the total number of created
electrons is
\begin{equation}
  \label{eqn:c}
  \mathcal{N} = \sum_{m,n} \norm{\Gud+-_{n;m}}^2 \,.
\end{equation}
Note that the quantity $\norm{\Gud+-_{n;m}}^2$ denotes the probability
that a negative-energy state with the quantum number $m$ turns into a
positive-energy state with the quantum number $n$.  A similar
calculation yields the expectation value of the total number of created
positrons $\sum_{m,n} \norm{\Gud-+_{n;m}}^2$.  In order to keep the
manuscript compact, we will present results for the created electrons
only.  Due to the symmetry of the pair creation process, corresponding
results for the positron can be deduced from calculations for the
electron.

\section{Numerical solution of the Dirac equation in monochromatic fields}
\label{sec:numerics}

In Sec.~\ref{sec:theory} we outlined that the number of created pairs
can be determined via the quantities $\Gud\pm\pm_{n;m}$, which can also
be written as
\begin{equation}
  \Gud\pm\pm_{n;m} =
  \int \herm{\phiout_n(\vec r)} 
  \phiin_m(\vec r, \tout)\,\D^3r \,,
\end{equation}
where $\phiin_m(\vec r, t)$ is a solution of the time-dependent Dirac
equation \eqref{eqn:Dirac_eq} with the initial condition $\phiin_m(\vec
r, \tin)=\phiin_m(\vec r)$. Thus, we can analyze the quantum
electrodynamical problem of pair creation by solving a large number of
independent time-dependent single-particle problems for the Dirac
equation \eqref{eqn:Dirac_eq}.  For the setup of Sec.~\ref{sec:beams}, it
is appropriate to solve these in momentum space.  The (real-valued)
total vector potential $\vec{A}=\vec{A}_{+}+\vec{A}_{-}$ can always be
written in the form
\begin{equation}
  \label{eqn:A_complex}
  \vec A(\vec r, t) = 
  \vec a(t)\e^{-\I kz} + \vec a(t)^*\e^{\I kz}
\end{equation}
with some complex-valued vector $\vec a(t)$ and its conjugate $\vec
a(t)^*$.  Because the external laser field was assumed to be
monochromatic, the operator $\vec\alpha\cdot\vec A$ (and consequently
also the Dirac Hamiltonian) couples momentum eigenstates with momentum
$\vec p$ only to momentum eigenstates with momentum $\vec p\pm k\vec
e_z$, which simplifies the numerical treatment considerably.  Because of
the monochromatic field \eqref{eqn:A_complex}, the Hilbert space
separates into disjunct subspaces, and each of them is spanned by plane
waves with momentum $n k\vec e_z + \vec p_0$ with integer $n$, where the
base momentum $\vec p_0$ selects a specific subspace.  To take advantage
of this symmetry, we expand the single-particle wave functions into
plane waves as
\begin{equation}
  \label{eqn:Dirac_wave_function}
  \Psi(\vec r, t) = 
  \sum_{n,\gamma} d_n^\gamma(t)  
  \sqrt{\frac{k}{2\pi}}\psi^\gamma
  \e^{\I ((n k\vec e_z + \vec p_0)\cdot \vec r)}
\end{equation}

with $\gamma\in\{1,2,3,4\}$ and the vectors
\begin{subequations}
  \begin{align}
    \psi^1 & = \trans{(1, 0, 0, 0)}\,, \\
    \psi^2 & = \trans{(0, 1, 0, 0)}\,, \\
    \psi^3 & = \trans{(0, 0, 1, 0)}\,, \\
    \psi^4 & = \trans{(0, 0, 0, 1)}\,.
  \end{align}
\end{subequations}
The expansion \eqref{eqn:Dirac_wave_function} into a \emph{discrete} set
of plane waves would also be justified for multicolor setups provided
that all wave vectors are an integer multiple of $k\vec e_z$.  Defining
\begin{equation}
  d_n(t)=\trans{(d_n^{1}(t),d_n^{2}(t),d_n^{3}(t),d_n^{4}(t))}
\end{equation}
and using the expansion \eqref{eqn:Dirac_wave_function}, the Dirac equation
reads
\begin{multline}
  \label{eqn:Dirac_equation_momentum-space}
  \I \dot d_n(t) = 
  (\vec\alpha\cdot (n k\vec e_z + \vec p_0) + \beta m_0 ) d_n(t) \\
  -w(t)q\vec\alpha\cdot \vec a(t) d_{n-1}(t) 
  -w(t)q\vec\alpha\cdot \vec a(t)^* d_{n+1}(t)
\end{multline}
in momentum space. Here, the window function
\begin{equation}
  \label{eqn:turn_on_and_off}
  w(t) = 
  \begin{cases}
    \sin^2 \frac{\pi (\Delta T-t)}{2\Delta T} & \text{if $-\Delta T\le t\le 0$,} \\
    1 & \text{if $0\le t\le T$,} \\
    \sin^2 \frac{\pi (T+\Delta T-t)}{2\Delta T} & \text{if $T\le t\le T+\Delta T$,} 
  \end{cases}
\end{equation}
has been introduced to model a smooth turn on and turn off of the
external laser field at $t=\tin=-\Delta T$ and $t=\tout=T+\Delta T$.

As the vector potential vanishes at $\tin$ and $\tout$, the states
$\phiin_n(\vec r)$ and $\phiout_n(\vec r)$ are free-particle states with
definite momentum, energy, and spin orientation (in the $z$~direction),
which read explicitly
\begin{subequations}
  \begin{align}
    \phiin[+]_{n,\uparrow/\downarrow}(\vec r) =
    \phiout[+]_{n,\uparrow/\downarrow}(\vec r) & = 
    u_{n k\vec e_z + \vec p_0,\uparrow/\downarrow}
    \e^{\I(n k\vec e_z + \vec p_0)\cdot\vec r}
    \,, \\
    \phiin[-]_{n,\uparrow/\downarrow}(\vec r) =
    \phiout[-]_{n,\uparrow/\downarrow}(\vec r) & = 
    v_{n k\vec e_z + \vec p_0,\uparrow/\downarrow}
    \e^{\I(n k\vec e_z + \vec p_0)\cdot\vec r} \,,
  \end{align}
\end{subequations}
where we have replaced the generic quantum number $n$ by the quantum
numbers $(n,\uparrow)$ or $(n,\downarrow)$, which specify the particle's
momentum ($n k\vec e_z + \vec p_0$) and its spin state (up or down).  We
also introduced
\begin{subequations}
  \begin{align}
    \label{eqn:pos_mom_spinor}
    u_{\vec p,\uparrow/\downarrow} & = 
    \sqrt{\frac{m_0 + \mathcal{E}(\vec p)}{2\mathcal{E}(\vec p)}}
    \begin{pmatrix}
      \vec\chi_{\uparrow/\downarrow}\\[1ex] 
      \dfrac{\vec\sigma\cdot\vec p}{m_0 + \mathcal{E}(\vec p)}
      \vec\chi_{\uparrow/\downarrow}
    \end{pmatrix}\,, \\
    \label{eqn:neg_mom_spinor}
    v_{\vec p,\uparrow/\downarrow}  & = 
    \sqrt{\frac{m_0 + \mathcal{E}(\vec p)}{2\mathcal{E}(\vec p)}}
    \begin{pmatrix}
      -\dfrac{\vec\sigma\cdot\vec p}{m_0 + \mathcal{E}(\vec p)} \vec\chi_{\uparrow/\downarrow} \\
      \vec\chi_{\uparrow/\downarrow}
    \end{pmatrix}
  \end{align}
\end{subequations}
with $\vec\chi_{\uparrow}=\trans{(1, 0)}$ and
$\vec\chi_{\downarrow}=\trans{(0, 1)}$ and $\mathcal{E}(\vec
p)=\sqrt{\vphantom{m_0}\smash[b]{m_0^2 +\vec p^2}}$.  In Fourier space
the states $\phiin[+]_{n,\uparrow/\downarrow}(\vec r)$ and
$\phiout[+]_{n,\uparrow/\downarrow}(\vec r)$ correspond to
\begin{subequations}
  \begin{equation}
    d_l =
    \begin{cases}
      u_{n k\vec e_z + \vec p_0, \uparrow/\downarrow} & \text{for $l=n$} \\
      \trans{(0, 0, 0, 0)} & \text{else,}
    \end{cases}
  \end{equation}
  while $\phiin[-]_{n,\uparrow/\downarrow}(\vec r)$ and
  $\phiout[-]_{n,\uparrow/\downarrow}(\vec r)$ correspond to
  \begin{equation}
    d_l =
    \begin{cases}
      v_{n k\vec e_z + \vec p_0, \uparrow/\downarrow} & \text{for $l=n$} \\
      \trans{(0, 0, 0, 0)} & \text{else.}
    \end{cases}
  \end{equation}
\end{subequations}
Choosing $\phiin[-]_{m,\uparrow}(\vec r)$ as the initial state at
$t=\tin=-\Delta t$, the desired transition matrix element for
transitions from the negative-energy state $(m,\uparrow)$ to the
positive-energy state $(n, \uparrow)$ reads in terms of the Fourier
space coefficients
\begin{equation}
  \Gud+-_{n, \uparrow;m, \uparrow} = 
  \herm{u_{n k\vec e_z + \vec p_0,\uparrow}} d_m(T)
\end{equation}
and analogously for other transitions.  The average number of created
electrons with momentum $n k\vec e_z + \vec p_0$ and spin up/down is
given by
\begin{equation}
  \label{eqn:cn_up_down}
  \mathcal{N}_{n, \uparrow/\downarrow} = 
  \sum_m 
  | \Gud+-_{n,\uparrow/\downarrow;m,\uparrow} |^2 +
  | \Gud+-_{n,\uparrow/\downarrow;m,\downarrow} |^2 \,.
\end{equation}
Thus, the spin-resolved average total number of created electrons
becomes
\begin{equation}
  \mathcal{N}_{\uparrow/\downarrow} = 
  \sum_{n} \mathcal{N}_{n, \uparrow/\downarrow} \,,
\end{equation}
and the average total number of created electrons equals
\begin{equation}
    \mathcal{N} = 
    \mathcal{N}_{\uparrow} + \mathcal{N}_{\downarrow} \,.
\end{equation}

The Dirac equation in momentum space
\eqref{eqn:Dirac_equation_momentum-space} is an infinite system of
ordinary differential equations.  In a numerical solution, however, we
have to limit the system to a finite number of modes.  This is justified
because the interaction term $w(t)q\vec\alpha\cdot\vec a(t)$ does not
depend on the mode number $n$, but the free Dirac Hamiltonian grows
asymptotically linearly with $n$.  Thus, for large $n$ transitions
between different free-particle states are suppressed, and consequently
large-$n$ states do not contribute to pair creation.  The system of
ordinary differential equations
\eqref{eqn:Dirac_equation_momentum-space} is solved via a Fourier split
operator method similar to the method presented in
Ref.~\cite{Bauke:Keitel:2011:Accelerating_Fourier}.

\section{Numerical results}
\label{sec:results}

\subsection{Single-mode dynamics}
\label{sec:example}

\begin{figure}
  \centering
  \includegraphics[scale=0.4]{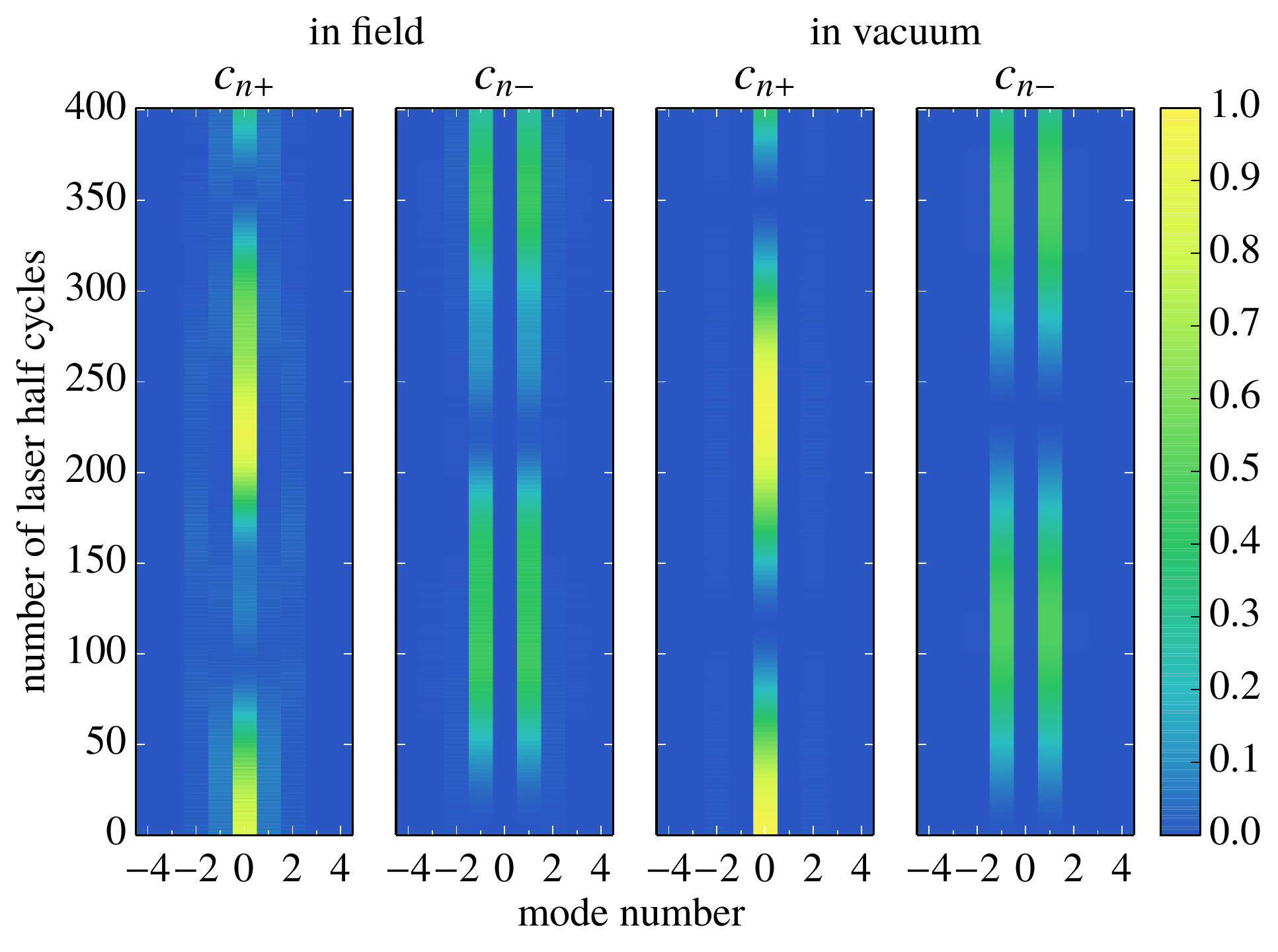}
  \caption{(color online) Time evolution of the initial momentum
    eigenstate $u_{\vec{p}=0,\uparrow}$ and its spreading into
    neighboring positive and negative momentum eigenstates for $\alpha_-
    = \alpha_+ = \pi/4$, $\omega=0.472\,m_0$ and $\xi = 1/2$. The
    vertical axis specifies the interaction time with the external
    field, and the horizontal axis specifies the momentum mode $\vec{p}
    = n k \vec{e}_z + 0$.  In field: The propagated state is decomposed
    into free momentum eigenstates without turning off the laser
    field. This is not unambiguous for physical interpretation. In
    vacuum: The external field has been turned off smoothly before
    decomposing the propagated wave function, allowing for a physical
    interpretation as momentum eigenstates.}
  \label{fig:mode_evo}
\end{figure}
\begin{figure}
  \centering
  \includegraphics[scale=0.4]{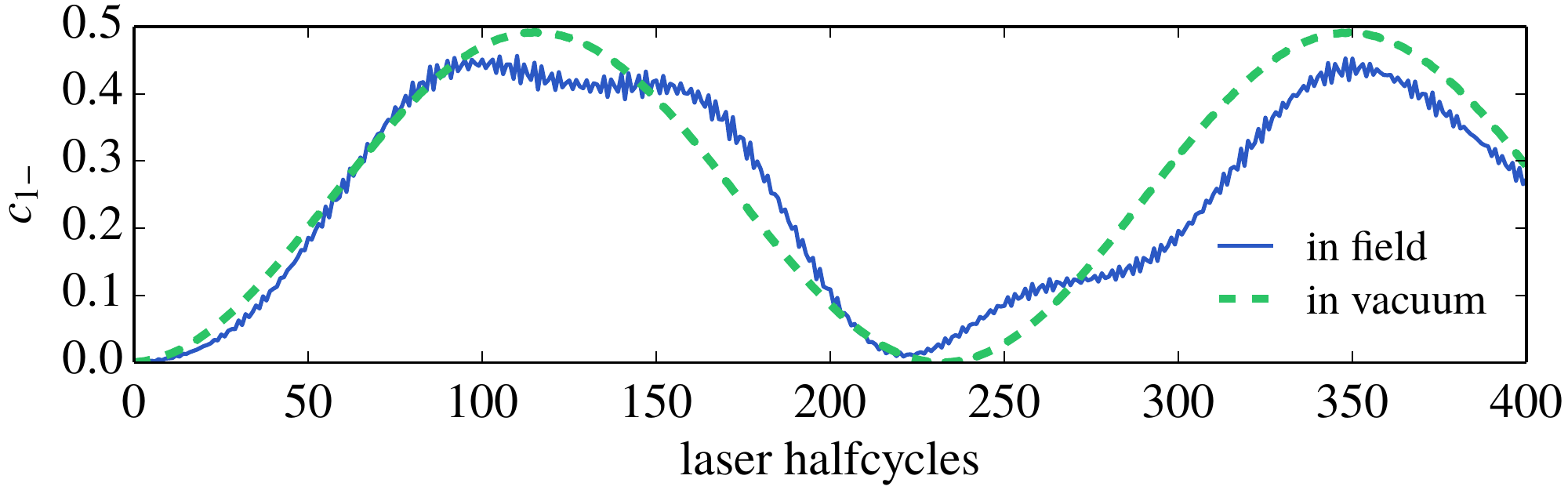}
  \caption{(color online) More detailed plot of the difference between
    taking an expectation value, while the electromagnetic field is
    still acting (in field) or has been smoothly turned off (in
    vacuum). Both lines correspond to the plotted weights $c_{n-}$
    (\ref{eqn:mode_weight_neg}) of Fig.~\ref{fig:mode_evo} with mode
    number~1.  Taking the expectation values in the field leads to no
    unique physical identification and also shows faster oscillations on
    the order of the laser frequency $\omega$. In contrast, a
    measurement after the field has been turned off yields a clear and
    smooth Rabi oscillation with respect to the interaction time with
    the external laser.}
  \label{fig:mode_diff}
\end{figure}

To help the reader understanding the following simulations, we would
like to start with an example of a single time evolution. Both lasers
are set up with linear polarization, i.\,e., $\alpha_- = \alpha_+ =
\pi/4$, and $\xi = 1/2$, where $\xi$ is determined with respect to a
single laser field.  Taking into account the electromagnetic fields of
both lasers would yield larger values of $\xi$, in particular $\xi = 1$
in case of linear polarization.  To calculate the expectation value of
an electron being created with zero momentum and spin up, the initial
state must be set to $u_{\vec{p} = 0, \uparrow}$; see
Eq.~\eqref{eqn:pos_mom_spinor}.  Now, according to Eq.~\eqref{eqn:cn},
the propagation back in time needs to be calculated. The duration of the
external electromagnetic field can be varied and will be given by the
number of half-cycles of the lasers period
\begin{equation}
  T = \text{(number of half-cycles)} \times
  \frac{\pi}{\omega} \,.
  \label{eqn:laser_duration}
\end{equation}
During this evolution (including the turn on \footnote{Due to the fact
  that the propagation is back in time, the mentioned ``turn on'' really
  corresponds to the turn-off, as the propagation starts at the end of
  the laser field. However, a positive time flow seems more natural for
  the discussion of the numerical results and therefore we stick to the
  labeling ``turn on''}), the initially localized wave function spreads
in momentum space to neighboring ``modes'' with momentum $\vec{p} = n k
\vec{e}_z + \vec{p}_0$. Furthermore, the wave function will also start
to occupy negative energy modes $v_{\vec{p},\uparrow/\downarrow}$.  Let
$\varphi(t)$ be the propagated wave function; then
\begin{subequations}
  \begin{align}
    \label{eqn:mode_weight_pos}
    c_{n+}(t) & = \norm{\scalprod{u_{\vec{p},\uparrow}}{\varphi(t)}}^2
    + \norm{\scalprod{u_{\vec{p},\downarrow}}{\varphi(t)}}^2 \,, \\
    \label{eqn:mode_weight_neg}
    c_{n-}(t) & = \norm{\scalprod{v_{\vec{p},\uparrow}}{\varphi(t)}}^2
    + \norm{\scalprod{v_{\vec{p},\downarrow}}{\varphi(t)}}^2 
  \end{align}
\end{subequations}
define the weights (probability coefficients) of this wave function to
be in a positive ($c_{n+}$) or negative ($c_{n-}$) energy mode with
momentum $\vec{p} = n k \vec{e}_z + \vec{p}_0$.  Here, the spin degree
of freedom has already been summed out. This decomposition into free
momentum eigenstates is not suited for physical interpretation
\cite{mocken2010nonperturbative} if the external field is still acting,
because the free momentum eigenstates do not represent eigenstates of
the Dirac Hamiltonian in the presence of an external electromagnetic
field. Hence, for a unique physical interpretation, the external field
is turned off smoothly, and then the decomposition into free momentum
eigenstates is carried out. Both cases, with and without turn off, are
shown in Fig.~\ref{fig:mode_evo}.  They look quite the same for this set
of parameters, but indeed they are different. This difference is
visualized in more in detail in Fig.~\ref{fig:mode_diff} for the weight
$c_{n-}$ with mode number $n=1$.  The measurement in the field contains
oscillation noise with frequencies on the order of the laser
frequency. These unphysical oscillations vanish, if the measurement is
performed in a properly switched-off field.  For the measurement in
vacuum, all plots in Figs.~\ref{fig:mode_evo}~and~\ref{fig:mode_diff}
show a slow oscillation with respect to the duration of the
electromagnetic field, corresponding to a Rabi oscillation of
$\mathcal{N}_{0,\uparrow}$, which is the expectation value for the
creation of an electron with zero momentum and spin up, given by the sum
over all weights $c_{n-}$ in the negative-energy subspace.  Similar Rabi
oscillations have been predicted for the relativistic Kapitza--Dirac
effect
\cite{Ahrens:Bauke:etal:2012:Spin_Dynamics,Ahrens:Bauke:etal:2013:Kapitza-Dirac_effect}.

\subsection{Rabi oscillations}
\label{sec:rabi_osc}

\begin{figure}
  \centering
  \includegraphics[scale=0.4]{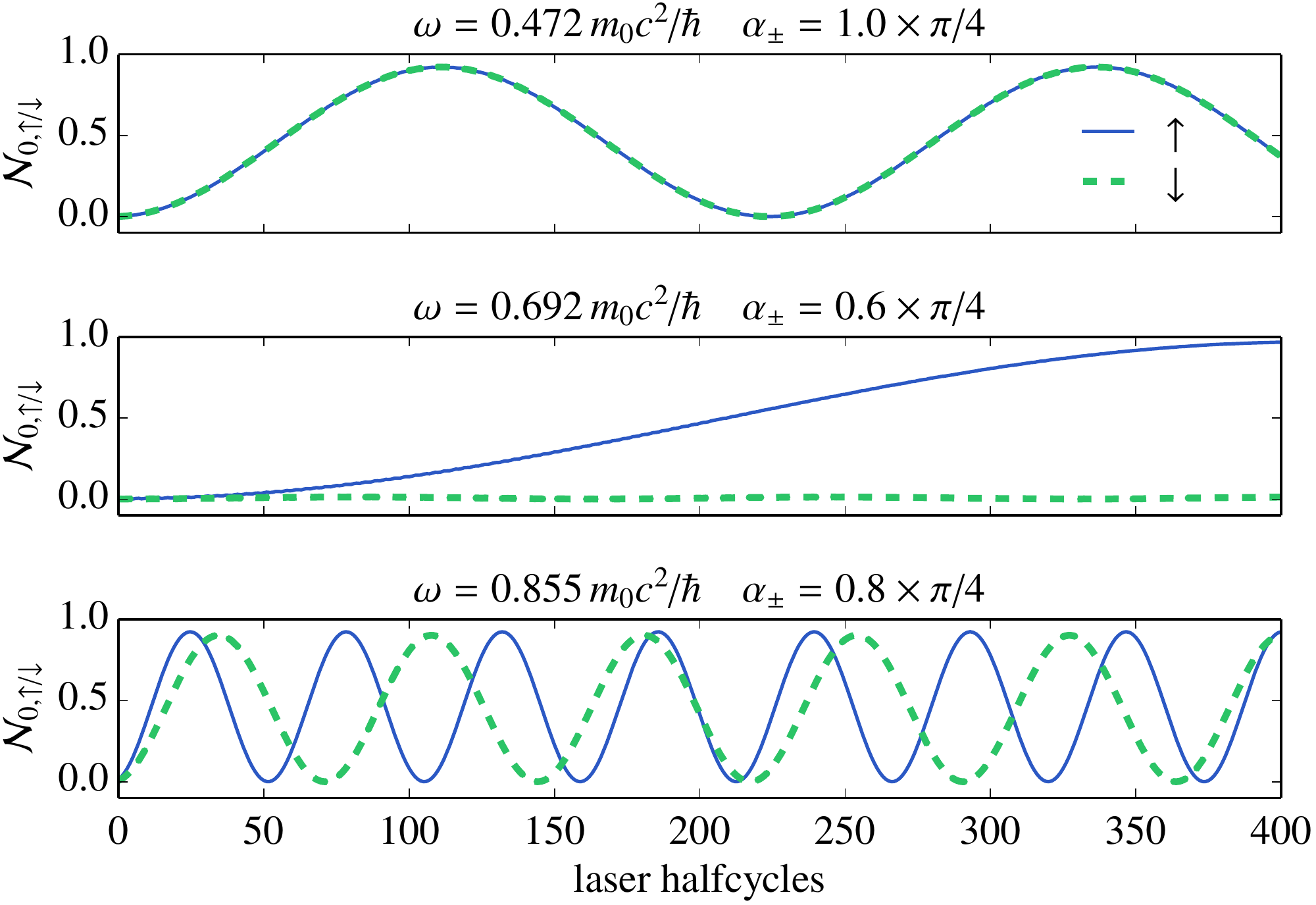}
  \caption{(color online) Rabi oscillations for the average occupation
    number $\mathcal{N}_{0,\uparrow/\downarrow}$ of electrons with final
    momentum $\vec{p} = 0$. Three different parameter sets for the
    polarization parameters $\alpha_-$ and $\alpha_+$ have been chosen,
    while $\xi = 1/2$ is kept for all. Top: The clearly visible resonant
    Rabi oscillation is independent of the final spin orientation due to
    the linear polarization.  Middle: Only the electron with spin up
    is at resonance for this specific elliptical polarization set, while
    the electron with spin down is completely off resonant.  Bottom: Both
    spin orientations are slightly off resonant with almost the same
    oscillation amplitude but different Rabi frequencies.}
  \label{fig:rabi_osc}
\end{figure}

It was mentioned in the previous subsection that the expectation values
for creating electrons and positrons with specific momentum and spin
(given by the occupation number \eqref{eqn:cn_up_down}) will exhibit
Rabi-type oscillations with respect to the interaction time in the
external field. Choosing a particular set of values for the polarization
parameters $\alpha_+$ and $\alpha_-$, the field strength (determined via
$\xi$), and the final momenta and spin $(\vec{p},
\uparrow\!/\!\downarrow)$ for the created particles of interest, pair
creation becomes resonant for specific frequencies of the laser field.
Three distinctive parameter sets are shown in Fig.~\ref{fig:rabi_osc}.
For all plots, the parameter $\xi$ equals $1/2$, and the final momenta of
interest for the created electron is set to $\vec{p} = 0$. The upper
plot shows that the resonance of the average number of created electrons
is independent on their final spin in a laser setup with linear
polarization.  For $\alpha_+ = \alpha_- = 0.6\times\pi/4$, however, a
spin effect can be observed as illustrated in the middle part of
Fig.~\ref{fig:rabi_osc}.  The created electron with spin up is at
resonance, while the electron with spin down is clearly off resonant and
does not show any noticeable Rabi oscillation.  The bottom plot shows
that for an elliptical polarization ($\alpha_+ = \alpha_- =
0.8\times\pi/4$, not far from linear polarization) both Rabi
oscillations of the two possible spin orientations are slightly off
resonant and almost equal in their amplitude, but their Rabi frequencies
differ.

\subsection{Resonance spectra}
\label{sec:spectra}

\begin{figure}
  \centering
  \includegraphics[scale=0.4]{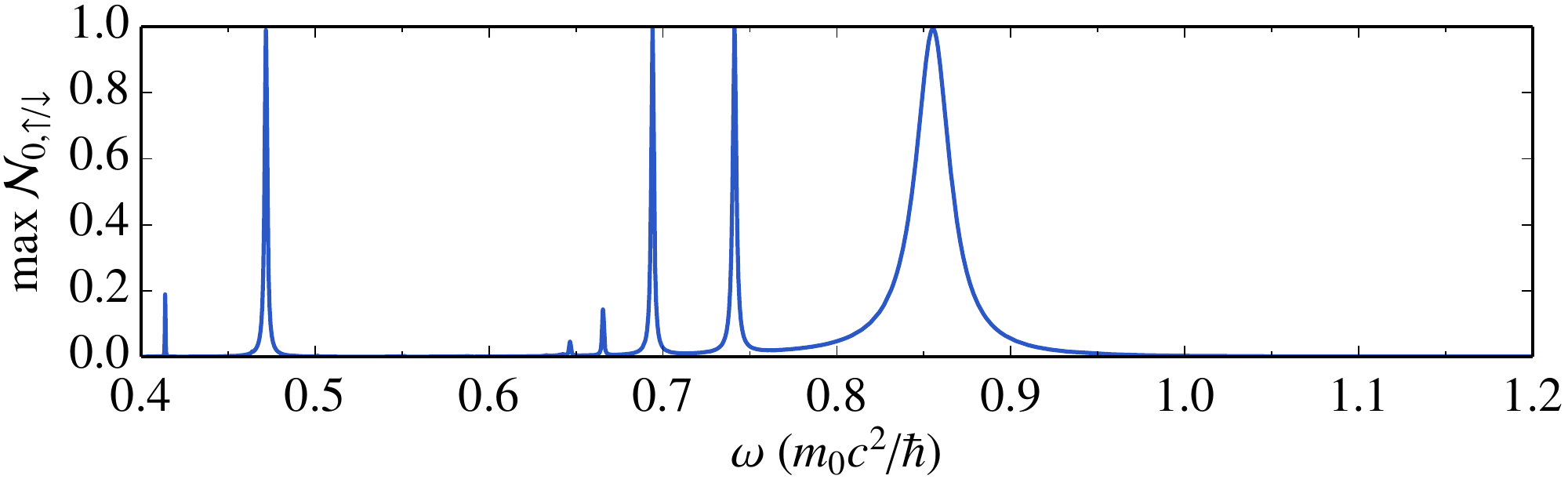}
  \caption{(color online) Pair production resonances for linear
    polarization $\alpha_- = \alpha_+ = \pi/4$ and $\xi = 1/2$. For each
    laser frequency $\omega$, a simulation of the type in
    Sec.~\ref{sec:rabi_osc} is performed, and the maximum value of
    $\mathcal{N}_{0,\uparrow/\downarrow}$ is taken as the ordinate value
    for this frequency. The spin of the produced electron needs not be
    distinguished for electrons with zero momentum due to the linear
    polarization. Four distinct resonance peaks can be seen with
    increasing resonance width as the laser frequency increases. The
    smaller peaks have a very low Rabi frequency and do not reach a full
    Rabi cycle as the maximum interaction time of this simulation is
    limited to 401 half-cycles of the laser field.  For longer
    interaction times, these resonance peaks will grow, and new ones may
    emerge.}
  \label{fig:lin_prob}
\end{figure}

\begin{figure}
  \centering
  \includegraphics[scale=0.4]{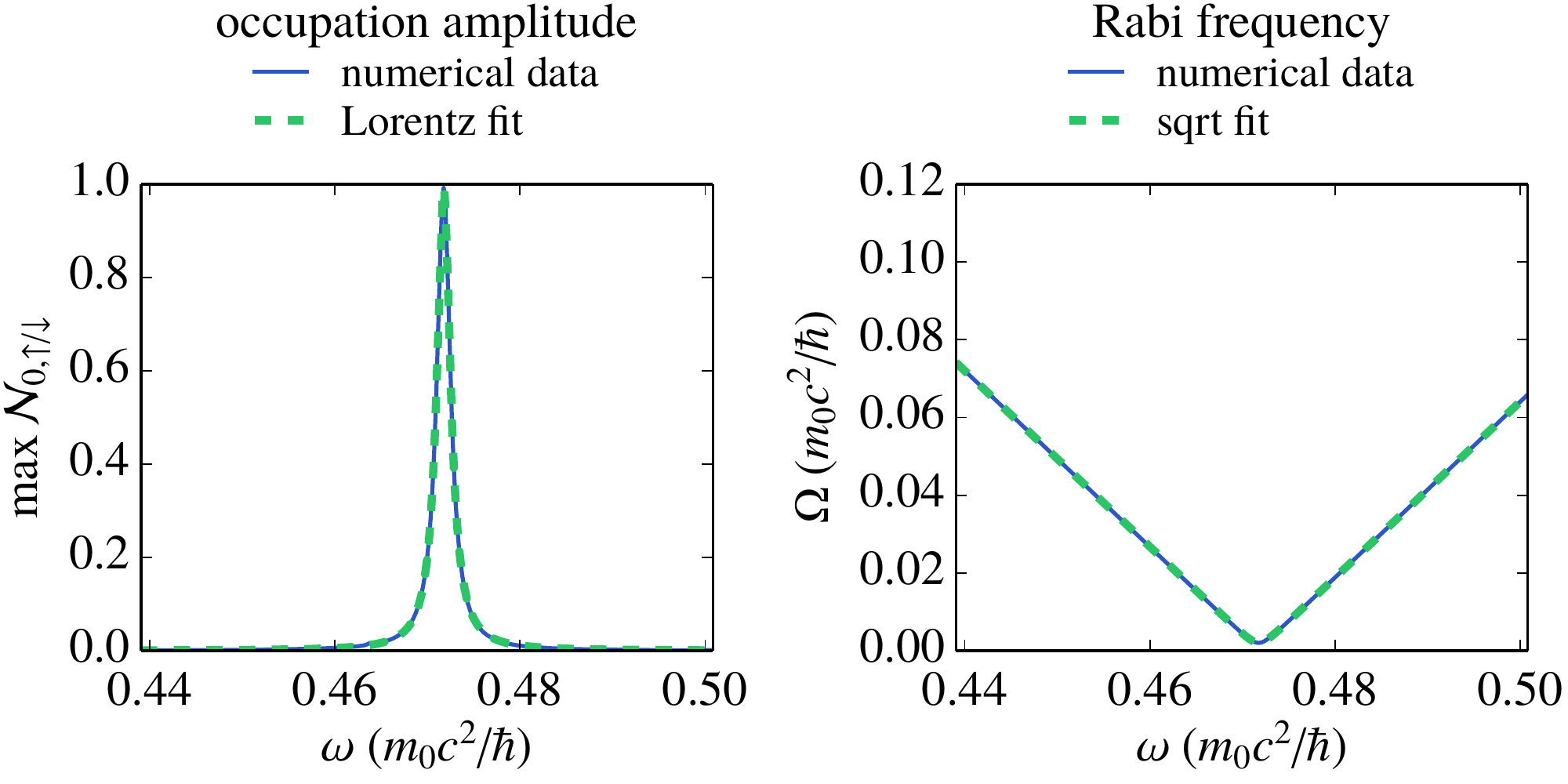}
  \caption{(color online) Resonance at $\omega \approx 0.472\,m_0$ for
    linearly polarized waves with $\alpha_+ = \alpha_- = \pi/4$ and $\xi
    = 1/2$ in detail.  Left: The pair production amplitude resonance
    including a Lorentz fit (\ref{eqn:lor_shape}) for the resonance
    peak. Right: The Rabi oscillation frequency $\Omega$ and its
    dependence on the detuning, including a fit of the form
    (\ref{eqn:sqrt_shape}).  Both fits agree perfectly well. The
    numerical fit values for the Lorentz fit are $\omega_R =
    0.47174(1)\,m_0$ and $w = 0.001780(3)\,m_0$, and a fit to
    \eqref{eqn:sqrt_shape} yields $\Omega_R = 0.002012(4)\,m_0$, $\eta =
    2.26639(4)$, and $\omega_R = 0.47175(1)\,m_0$.  }
  \label{fig:det_res}
\end{figure}

\begin{figure*}
  \centering
  \includegraphics[scale=0.4]{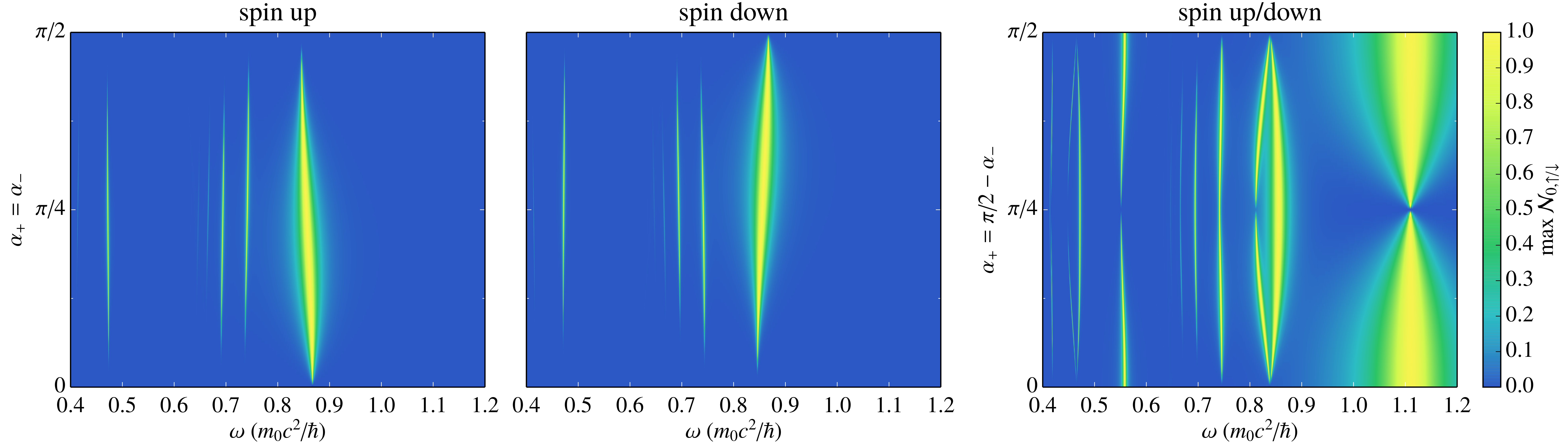}
  \caption{(color online) Polarization and laser frequency dependence of
    the electron creation probability with zero final momentum and
    specific spin orientation with respect to the positive
    $z$~direction. Both lasers field strengths' are determined by their
    frequency $\omega$ and $\xi=1/2$.  The color encodes the maximum
    electron creation probability measured for a time of maximum 400
    half-cycles of the laser fields period. Various resonances can be
    seen, which change their resonance frequency $\omega_R$ with varying
    ellipticity $\alpha_\pm$.
    Left two plots: Both plots correspond to the case of corotating
    laser fields.  A slice along the line of linear polarization
    $\alpha_+=\pi/4$ would correspond to the spectrum
    Fig.~\ref{fig:lin_prob}.  Except for $\alpha_+ = \pi/4$, which
    corresponds to linear polarization, electron creation probabilities
    depend on the spin orientation of the created electron.   Note that
    pair creation is completely suppressed for circular polarized light
    ($\alpha_\pm = 0$ or $\alpha_\pm = \pi/2$). 
    Right plot: This represents the resonance spectra for antirotating
    fields, where the electron creation probabilities do not depend on
    the spin orientation.}
  \label{fig:pol_prob}
\end{figure*}

In this and the following two sections, we will consider particles
having zero momentum only, while Sec.~\ref{sec:non-zero_momentum} will
allow for the case of nonzero momentum.  Having examined the Rabi
oscillations for specific laser frequencies, we will analyze the
pair-creation spectra now.  Each data point of a pair-creation spectrum
represents the maximum value of $\mathcal{N}_{0,\uparrow/\downarrow}$
for a specific set of laser parameters.  Figure~\ref{fig:lin_prob} shows
the maximum value of $\mathcal{N}_{0,\uparrow/\downarrow}$ as a function
of the laser frequency $\omega$ while keeping all other parameters
fixed.  Here the polarization is linear, i.\,e., $\alpha_- = \alpha_+ =
\pi/4$, and $\xi = 1/2$.  Note that in case of linear polarization
pair-creation probabilities do not depend on the spin orientation, and
therefore $\mathcal{N}_{0,\uparrow}=\mathcal{N}_{0,\downarrow}$.  Four
complete resonances can be inferred from Fig.~\ref{fig:lin_prob}. Their
resonance width increases with increasing frequency. Furthermore, three
smaller resonances appear that do not reach the full resonance maximum
of unity. The reason for this is that their Rabi oscillation frequencies
are very low and the top of the Rabi cycle has not been reached before
the simulation ends as the maximum interaction time with the external
field has been limited to 401 half-cycles of the external laser field
for this simulation.

A close-up sweep of the resonance at $\omega \sim 0.472\,m_0$ has been
done to show the typical Rabi characteristics. One characteristic, the
Lorentz type resonance shape, is plotted on the left of
Fig.~\ref{fig:det_res} with a fit to the Lorentz function
\begin{equation}
  \max\mathcal{N}_{0,\uparrow/\downarrow} = 
  \frac{1}{1 + \left( 
      \frac{\omega - \omega_R}{w/2} 
    \right)^2} \,.
  \label{eqn:lor_shape}
\end{equation}
The fit parameter $\omega_R$ corresponds to the resonance frequency, and
$w$ gives to the full width at half maximum. Furthermore, the Rabi
oscillation frequency $\Omega$ can be extracted from the numerical data
for each laser frequency $\omega$. The right part of
Fig.~\ref{fig:det_res} shows the Rabi frequency $\Omega$ together with a
fit to the function
\begin{equation}
  \Omega = \sqrt{\Omega_R^2 + \eta^2 (\omega - \omega_R)^2}\,,
  \label{eqn:sqrt_shape}
\end{equation}
where $\Omega_R$ denotes the Rabi frequency at resonance and $\eta$ is a
fit coefficient depending on the resonance. It can be seen that both
fits agree perfectly well with the numerical data and hence justify the
categorization of pair production in this parameter regime as a Rabi
process.


Extrapolating these results into the regime of frequencies lower than
shown in Fig.~\ref{fig:lin_prob}, the resonance peaks become very dense
with respect to the frequency axis.  This is due to the fact that the
positions of neighboring resonances, which differ in the number of
absorbed photons by 1, will differ only by the small photon energy
corresponding to the low laser frequencies. The Rabi frequencies of the
resonances will decrease for the increasing number of photons needed to
create a pair; see Ref.~\cite{mocken2010nonperturbative}. Hence, only
the very beginning of the Rabi oscillations and thus a quadratic time
dependence of the average occupation number for a created pair with
sharp momentum at a specific resonance will be observable.
Incorporating the finite resolution of measuring the momenta of the
produced pairs and the corresponding integration in momentum space
(analogous to Fermi's golden rule), we expect that this leads to a
linear time dependence, i.\,e., a rate for pair production, which is
characteristic for low laser frequencies in the tunneling regime.

\subsection{Polarization dependence of the resonance spectra}
\label{sec:polarization}

Going one step further, the resonance structure in
Fig.~\ref{fig:lin_prob} can also be calculated for different
polarizations than $\alpha_+ = \alpha_- = \pi/4$. Performing a
two-dimensional sweep over the laser frequency $\omega$ and the
polarization parameters $\alpha_+$ and $\alpha_-$ leads to the
two-dimensional color plots shown in Fig.~\ref{fig:pol_prob}.  The
polarization parameters $\alpha_+$ and $\alpha_-$ are not varied
independently; instead they either fulfill $\alpha_+ = \alpha_-$
(corotating fields) or $\alpha_- = \pi/2 - \alpha_+$ (antirotating
fields).

Figure~\ref{fig:pol_prob} shows the spin resolved electron creation
probability for electrons with zero final momentum as a function of the
laser frequency $\omega$ and the polarization.  The two plots on the
left correspond to the spin-up and spin-down spectra for corotating
fields, while the single plot on the right presents the spectrum for
antirotating fields.  Our numerical results show that the spectra depend
on the spin orientation for corotating and are identical for both spin
orientations in case of anti-rotating fields.  Because spin-down and
spin-up spectra are different for corotating fields, see left part of
Fig.~\ref{fig:pol_prob}, it is possible to have for some fixed $\omega$
and $\alpha_+$ zero electron creation probability for a specific spin
orientation and nonzero probability for the opposite orientation.  For
antirotating fields, however, the electron creation probability is
completely symmetric with respect to the spin orientation of the created
electron as shown in the right part of Fig.~\ref{fig:pol_prob}.  Note
that pair creation is suppressed for $\alpha_+=0$ and $\alpha_+=\pi/2$,
i.\,e., circular polarization, in the case of corotating fields.  For
antirotating fields, however, resonances exist also for $\alpha_+=0$
and $\alpha_+=\pi/2$.

A possible interpretation of the different behavior of corotating and
antirotating fields is that the standing electromagnetic wave carries
nonzero spin density for corotating fields but not for antirotating
fields.  The orientation of the photonic spin density breaks the
symmetry with respect to reflection along the propagation axis of the
laser fields; see Fig.~\ref{fig:laser_spin}.  Consequently, pair
creation probabilities depend on the spin orientation of the created
electrons.  Note that for $\alpha_+=\pi/4$ (linear polarization) the
photonic spin density vanishes also for corotating fields.
Consequently, there is no dependence on the electron spin; see left part
of Fig.~\ref{fig:pol_prob}.  Pair creation probability becomes
spin dependent, however, if $\alpha_+ \ne \pi/4$.

\begin{figure*}
  \centering
  \includegraphics[scale=0.4]{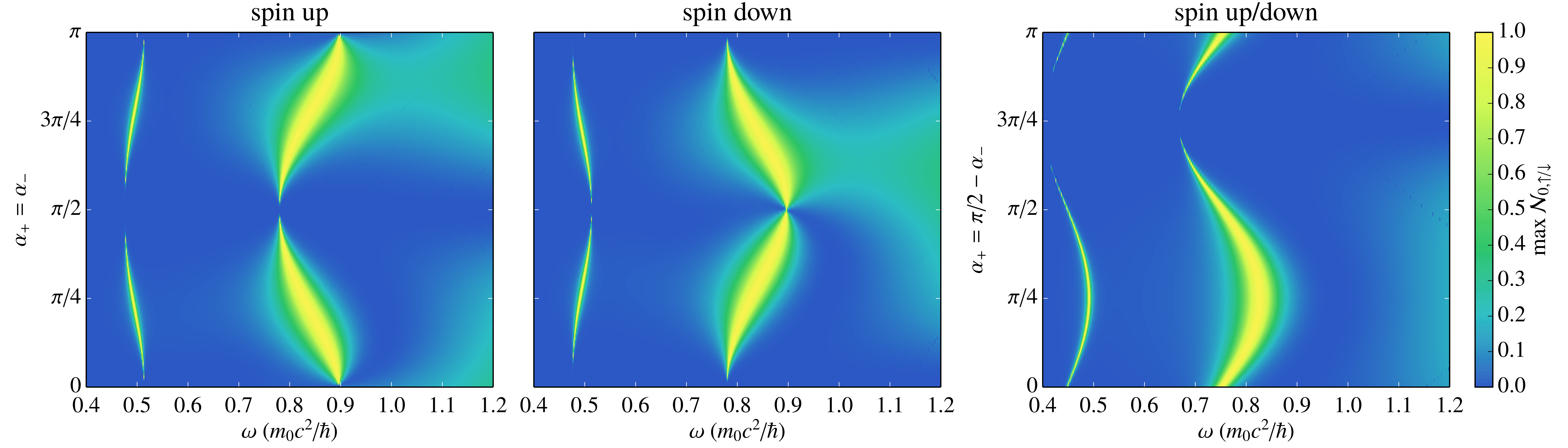}
  \caption{(color online) Polarization and laser frequency dependence of
    the electron creation probability with zero final momentum and
    specific spin orientation with respect to the positive $z$~direction
    when the dipole approximation is applied.  All parameters as in
    Fig.~\ref{fig:pol_prob}.}
  \label{fig:rotating_el_field}
\end{figure*}

The parameter $\alpha_+$ is restricted to the range from 0~to~$\pi/2$ in
Fig.~\ref{fig:pol_prob}.  One can show by the symmetry properties of the
standing electromagnetic wave's vector potential (or its electromagnetic
field) that for $\alpha_+ = \alpha_-$ the spectra in
Fig.~\ref{fig:pol_prob} must be periodic in $\alpha_+$ with period
$\pi$.  The part of the spectrum for $\pi/2\le\alpha_+\le\pi$, which is
not shown in Fig.~\ref{fig:pol_prob}, can be obtained by mirroring the
spectrum for $0\le\alpha_+\le\pi/2$ at $\alpha_+=\pi/2$.  Furthermore,
the spectrum is periodic in $\alpha_+$ with period $\pi/2$ for
antirotating fields.

Figure~\ref{fig:pol_prob} reveals an explicit spin dependence of the
resonance structures in case of corotating fields. These differences
become more pronounced the more the fields' ellipticity differs form
linear polarization at $\alpha_+ = \pi/4$.  The spectra for spin up, if
mirrored along the line $\alpha_+ = \pi/4$, are equal to the spectra for
spin down and vice versa. This can be understood by symmetry, because
flipping the spin of the final particle leads to the same result as
flipping the sense of rotation for the lasers' polarization, i.\,e.,
from right circulating polarization to left circulating
polarization. Looking at the resonance between $0.8\,m_0 < \omega <
0.9\,m_0$, for example, the resonance frequency $\omega_R$ decreases
toward circular right polarization $\alpha_+ \rightarrow \pi/2$ for spin
up, while it increases for spin down.

Note that this shift of the resonance frequency is always opposite for
electrons with spin up and with spin down.  This might be understood by
the coupling between the particle's spin and the lasers' magnetic field,
as seen in the rest frame of the particle.  With the electric and
magnetic field components $\vec{E}$ and $\vec{B}$ in the laboratory
frame, the magnetic field in the rest frame is in leading order $\vec B'
\sim \vec B - \vec{v} \times \vec{E}$.  Because $\vec{B}$ is always
perpendicular to the $z$~axis, the Zeeman energy $-q/(2m_0)\,
\vec{\sigma} \cdot\vec{B}'= q/(2m_0)\,\vec{\sigma}\cdot(\vec{v} \times
\vec{E})$ results, where $\vec{v}$ denotes the particle's velocity in
the laboratory frame.  For corotating laser fields with elliptical
polarization, the electric field rotates in the \mbox{$x$-$y$~plane} and
therefore leads to a two-dimensional trajectory of the particle in the
\mbox{$x$-$y$~plane}.  Because $\vec{E}$ also lies in the
\mbox{$x$-$y$~plane}, the magnetic field $\vec{v} \times \vec{E}$ points
in the $z$~direction yielding a modification of the particle's energy
and in this way modifying the resonance frequency $\omega_R$.  The sign
of this energy shift depends on the particle's spin in the
$z$~direction, which explains the opposite shift of the resonance
frequency for different spin orientations.  For antirotating fields, in
contrast, the electric field does not rotate in the
\mbox{$x$-$y$~plane}.  It oscillates with a fixed direction in the
\mbox{$x$-$y$~plane}.  Consequently, the particle's trajectory projected
onto the \mbox{$x$-$y$~plane} will always be parallel to the electric
field direction.  Thus, the Zeeman energy is zero for an electron with
spin in $z$~direction, and consequently there is no spin-dependent shift
of the resonance frequency.

The shift of the Zeeman energy, which we find in the case of corotating
fields, is also given by the nonrelativistic corrections to the Pauli
equation.  These correction can be obtained by an expansion of the Dirac
equation via a Foldy--Wouthuysen transformation and contains the term
\mbox{$1/(4m_0^2)\,\vec\sigma\cdot (\vec E\times (-\I\nabla-q\vec A))$}.
For zero canonical momentum, this is up to a factor of $2$ equivalent to
the shift of the Zeeman energy discussed above.  Note that the factor
$\vec E\times \vec A$ can be attributed to the photonic spin density of
the electromagnetic field
\cite{Bauke:Ahrens:etal:2014:Electron-spin_dynamics}, which is zero for
antirotating fields.

\subsection{Comparison to the dipole approximation}
\label{sec:dipole_approx}

When pair creation in standing electromagnetic waves is studied,
neglecting the magnetic field component and the spatial variation of the
electric field is a commonly applied approximation
\cite{Fedotov:Gelfer:etal:2011:Kinetic_equation,Blinne:Gies:2014:Pair_production,Li:etal:2014:Mass_shift}.
Therefore, we will analyze the effect of this approximation in this
subsection.  Neglecting the spatial variation of the electromagnetic
field given by the potential \eqref{eqn:vector_pot} yields the dipole
approximation vector potential
\begin{equation}
  \label{eqn:vector_pot_DA}
  \vec{A}_{\pm}(t) = \Re{\left(
      \frac{E}{\I\omega}
      \left( 
        \cos \alpha_\pm \ket{l} + \sin \alpha_\pm \, \e^{\I \varphi_\pm} \ket{r} 
      \right)
      \e^{-\I\omega t}
    \right)} \,.
\end{equation}
This dipole approximation of the electromagnetic fields is employed in
Fig.~\ref{fig:rotating_el_field} for comparison with
Fig.~\ref{fig:pol_prob}, where the full electromagnetic field was taken
into account.  Also in the framework of the dipole approximation, Rabi
oscillations of the electron creation probability and resonances can be
found.  The dipole approximation gives, however, results, which differ
quantitatively as well as qualitatively from the case of pair creation
in electromagnetic fields.  For example, the number of resonances is
reduced in case of the dipole approximation.  In
Fig.~\ref{fig:rotating_el_field} the parameter $\alpha_+$ ranges from
$0$~to~$\pi$ in order to point at differences to the simulation with the
full monochromatic field and to show some peculiarities of the turn on
and off of the electromagnetic field.

\begin{figure*}
  \centering
  \includegraphics[scale=0.4]{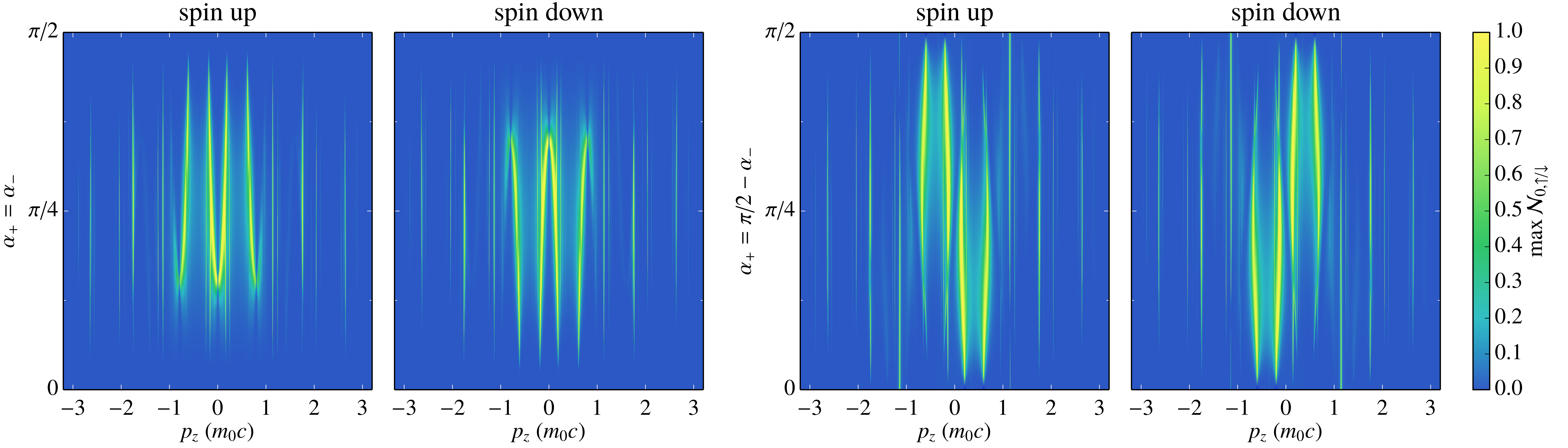}
  \caption{(color online) Polarization dependence of the pair creation
    probability, where the electron has different final momenta
    $\vec{p}_0 = (0, 0, p_z)$.  The lasers' frequencies are set to
    $\omega = 0.8\,m_0$, and their strength is implicitly determined by
    $\xi = 1/2$.  From all four plots can be inferred that, for this
    fixed frequency and strength, final pairs with distinguished momenta 
    will be produced.
    Left two plots: These two plots correspond to the
    corotating field setup and differ by the spin orientation of the final
    electron.  The vertical cut along $p_z = 0$ corresponds to a vertical cut
    along $\omega = 0.8\,m_0$ in Fig.~\ref{fig:pol_prob}.  Both plots are
    symmetric around $p_z = 0$. This is because inverting $p_z$ would also
    invert the final helicity of the created particles, but due to the fact
    that for corotating fields both fields' helicities are opposite to each
    other, the systems stays invariant. 
    Right two plots: These two plots correspond to the antirotating
    field setup. In contrast to the case of corotating fields, the
    spectra are not invariant under inversion of the $p_z$ momentum and
    consequently the helicity of the final particle, because both laser
    fields have the same helicity for this case. But if, in addition to
    the momentum inversion, also the spin is inverted, the helicity of
    the final particle does not change, and the spectrum stays the
    same.}
  \label{fig:single_mom}
\end{figure*}

In case of corotating fields, the two plots on the left of
Fig.~\ref{fig:rotating_el_field} are not perfectly symmetric with
respect to the line $\alpha_+ = \pi/2$, especially for $0.9\, m_0
\lesssim \omega \lesssim 1.2\, m_0$.  This is due to the turn on and
turn off as the following argument illustrates. Comparing, for example,
$\alpha_+ = \alpha_- = \pi/4$ with $\alpha_+ = \alpha_- = 3 \pi/4$, both
laser fields are linearly polarized with the electric field component
pointing in the $x$~direction and in the $y$~direction, respectively.
Due to rotational invariance around the $z$~axis, both parameters should
yield the same results.  But owing to the parametrization and analytical
structure of the vector potential (\ref{eqn:vector_pot_DA}), the
electric field amplitude is at maximum for $\alpha_+ = \alpha_- = \pi/4$
at $t=0$, while it is zero for $\alpha_+ = \alpha_- = 3 \pi/4$ at $t=0$
maximal. This phase shift leads in combination with the window function
(\ref{eqn:turn_on_and_off}) to an asymmetry, which has no effect in
regimes with Rabi frequencies much slower than the laser frequency,
because the pair production effect builds up over many laser cycles and
is less sensitive to the initial phase of the laser field.  For Rabi
frequencies on the order of the laser frequency as observed for $\omega
\gtrsim m_0$ this difference, however, becomes noticeable.

For the case of antirotating electromagnetic fields, we found in
Sec.~\ref{sec:polarization} that the probability to create an electron
with zero momentum is periodic in the field parameter $\alpha_+$ with
period $\pi/2$.  Applying the dipole approximation, however, the
spectrum becomes periodic in $\alpha_+$ with period $\pi$ as shown in
the right part of Fig.~\ref{fig:rotating_el_field}.  This is a direct
consequence of the symmetry properties of the vector potentials in
(\ref{eqn:vector_pot_DA}).  For $\alpha_+ = \pi/4$ there are broad
resonances, whereas pair creation is inhibited for $\alpha_+ = 3\pi/4$.
Both parameters correspond to electric fields with linear polarization.
For $\alpha_+ = \pi/4$ the vector potentials $\vec{A}_{+}(t)$ and
$\vec{A}_{-}(t)$ given in (\ref{eqn:vector_pot_DA}) superimpose
constructively, while both potentials cancel for $\alpha_+ = 3\pi/4$.
Note that this cancellation is a peculiarity of the dipole
approximation.  Two counterpropagating electromagnetic waves with
temporal and spatial variation as specified in (\ref{eqn:vector_pot})
cannot interfere such that both annihilate each other completely.

Notably, the simulation for the dipole approximation also yields a
strong dependence of the pair creation on the spin of the created
electron and the polarization of the electric field for corotating
fields while not for antirotating fields.  Comparing the magnitude of
the shift of the resonance frequency $\omega_R$ with the one in
Fig.~\ref{fig:pol_prob} shows that this effect is even stronger in the
dipole approximation with a rotating electric field only.  As mentioned
in the previous subsection, this shift might be due to the coupling of
the particles spin and the magnetic field, as seen in the rest frame of
the particle.  Alternatively, the spin dependence could also be inferred
from symmetry. In case of corotating electric fields, the sense of
rotation changes under reflection of the $z$~axis, and thus the pair
creation probability might become spin dependent.  In contrast, two
antirotating electric fields form an electric field with a constant
direction, and therefore the field configuration is symmetric under
reflection of the $z$~axis.  Consequently, also pair creation is
symmetric under reflection of the final particle's spin, assuming zero
final momentum.

Comparing the various resonances in Figs.~\ref{fig:pol_prob}
and~\ref{fig:rotating_el_field} in terms of the polarization dependent
shift of the resonance frequencies shows that the frequency shift
becomes smaller at lower laser frequencies.  The reason for this might
be a combination of the decreased amount of photon energy and an
averaging effect due to the increased number of photons.  The
polarization dependence of the resonance frequencies tends to be larger
for the dipole approximation than for the case of electromagnetic
fields, which may be attributed to the fact that the electric field is
maximal at each point in space within the dipole approximation and to
the absence of the magnetic field in the laboratory frame.

\subsection{Final momentum dependence of the resonance spectra}
\label{sec:non-zero_momentum}

In the previous subsections, we investigated the creation of particles
with zero final momentum.  Looking, for example, at
Fig.~\ref{fig:pol_prob}, it shows that at $\omega \sim 0.8\,m_0$ no
particles with zero final momentum can be created for the given setup.
Particles with nonvanishing momentum, however, might be created.  A
sweep over the final momentum allows us to study the momentum dependence
of the pair production probability.  Due to the fact, that a
three-dimensional sweep over all final momenta is numerically quite
expensive and also hard to visualize, we vary only the momentum directed
along the $z$~axis, i.\,e., the lasers' propagation axis.  The
perpendicular momenta are set to zero.  Furthermore, we chose a fixed
laser frequency of $\omega = 0.8\,m_0$ and additionally scanned over the
ellipticity for corotating and antirotating field configurations as done
in Figs.~\ref{fig:pol_prob} and~\ref{fig:rotating_el_field}.  The
numerical result is shown in Fig.~\ref{fig:single_mom}.  The two plots
on the left correspond to corotating fields, while the two on the right
correspond to antirotating fields.  In agreement with our previous
calculations, Fig.~\ref{fig:single_mom} indicates that no particles are
created with zero final momentum for the fixed frequency of $\omega =
0.8\,m_0$.  Pair creation is possible for some nonvanishing momenta as
the resonances in Fig.~\ref{fig:single_mom} indicate.  A notable
peculiarity for the case of corotating fields can be seen in the left
part of Fig.~\ref{fig:single_mom}.  Tracing the two broader resonances
close to $p_z = 0$ for the spin up case from top to bottom, they
approach each other but do not join each other at $p_z = 0$ for
$\alpha_+ \sim \pi/8$.  Thus, pair creation is suppressed at $p_z=0$
exactly consistent with Fig.~\ref{fig:pol_prob}.

These resonances in Fig.~\ref{fig:single_mom} exhibit distinctive
symmetries.  For corotating fields, both plots for spin up and spin down
are symmetric around the line $p_z = 0$.  This symmetry is related to
the fact that inverting the $z$ momentum preserves the relative
orientation of the helicities of the created electron and the two
counterpropagating waves.  Inverting the final momentum inverts the
helicity of the electron.  Because both laser fields' helicities are
opposite to each other for corotating fields, inverting the
$z$~momentum or equally the helicity of the created electron does not
change the system.  The electron's helicity is parallel to one wave and
antiparallel to the other as before inversion of the $z$~momentum.

Both helicities of the laser fields point into one preferred direction
for antirotating fields.  Thus, inverting the $z$~momentum does not
leave the system invariant, and electrons are created preferably with
helicity parallel or antiparallel to the helicity of the electromagnetic
fields.  To preserve the relative orientation of the helicities of the
created electron and the two counterpropagating waves, one has to invert
the $z$ momentum and the waves' helicity, which corresponds to the
combined transformation $p_z\to p_z$ and $\alpha_+\to\pi/2-\alpha_+$,
and thus the spectra in the right part of Fig.~\ref{fig:single_mom} are
invariant under this transformation.

Both the corotating as well as the antirotating case have in common that
mirroring the spectrum for one spin orientation around the $\alpha_+ =
\pi/4$ line yields the spectrum for the other spin orientation.  This
symmetry holds for the spectra in Fig.~\ref{fig:pol_prob} as well as in
Fig.~\ref{fig:single_mom}.  Again, this is a consequence of the system's
invariance under flipping the created electron's spin/helicity and the
field's spin/helicity.

Let us compare the spectra for corotating fields with the ones for
antirotating fields in Fig.~\ref{fig:single_mom} at fixed values of the
ellipticity parameter $\alpha_+$.  In the case of corotating fields,
electrons with spin up or down are created with higher probability
depending on the ellipticity parameter $\alpha_+$; see the left part of
Fig.~\ref{fig:single_mom}.  Because of the symmetry properties described
in detail above, the electrons are created with equal probability having
positive or negative helicity; this means created electrons propagate in
the positive or negative $z$ direction with equal probability.  This
changes if we consider antirotating fields; see the right part of
Fig.~\ref{fig:single_mom}.  Here electron creation is spin symmetric.
This means spin-up and spin-down electrons are created with equal
probability.  However, electrons with positive or negative helicity are
created with higher probability depending on the ellipticity parameter
$\alpha_+$.  For an ellipticity of $\alpha_+\sim \pi/8$, for example,
electrons are created preferably with spin up, which propagate only
along the positive $z$ direction, and with spin down, which propagate
only along the negative $z$ direction.

\section{Conclusions} 
\label{sec:conclude}

We studied pair creation in strong electromagnetic fields in the
nonperturbative regime, i.\,e., $\xi \sim 1$.  For this purpose, we
derived a numerical framework for solving the time-dependent Dirac
equation for an electron in two counterpropagating monochromatic laser
fields with equal frequency.  This setup is quite generic, because also
other setups with two lasers having unequal frequencies or a colliding
angle different than $180^{\circ}$ could be incorporated by a proper
Lorentz transformation.  Furthermore, the polarization and the field
strength of each individual laser field could be chosen arbitrarily.
With this we overcame the so often applied dipole approximation for the
laser fields and included the electric field's space dependence as well
as the magnetic field's influence on pair production, leading to true
three-dimensional dynamics.  For numerical feasibility and in order to
keep the space of possible parameters small, high laser frequencies and
equal intensity and polarization of the counterpropagating fields were
assumed.  The numerical method, however, is also capable of treating
setups with unequal polarizations and intensities.  It can be easily
generalized to multicolor setups, provided that all frequencies are
low-order harmonics of some base frequency.

It was shown that switching off the external laser fields is essential
for a proper determination of the pair creation probabilities.  These
probabilities exhibit the characteristic features of Rabi oscillations
with the corresponding resonance frequencies and Rabi frequencies.  The
resonance structures and their dependencies on the final quantum numbers
of the produced pairs as well as on the polarization of the laser field
were investigated numerically in more detail.

A quite remarkable result is the observation that, by tuning the
polarization of the two laser beams, spin polarized pairs can be
created.  By using corotating laser fields, for example, the spins of
the electrons in the produced pairs may be aligned all the same along
the propagation axis of the lasers.  Antirotating fields, in contrast,
may yield electrons with spin-up propagating forward and electrons with
spin down propagating backward with respect to the lasers' propagation
axis.

\begin{acknowledgments}
  We are thankful for stimulating discussions with S.~Ahrens and
  C.~M{\"u}ller at the onset of this work.
\end{acknowledgments}

\bibliography{refs}

\end{document}